%% file: main.tex
\documentclass[conference]{IEEEtran}
\IEEEoverridecommandlockouts

\usepackage{cite}
\usepackage{amsmath,amssymb,amsfonts}
\usepackage{algorithmic}
\usepackage{graphicx}
\usepackage{textcomp}
\usepackage{xcolor}
\usepackage{cases/cases}
\usepackage{subfig}
\usepackage{relsize}

\usepackage{anyfontsize}

\usepackage{caption}
\DeclareCaptionFont{figure_size}{\fontsize{8}{8}\selectfont}
\DeclareCaptionFont{normal_size}{}

\def\BibTeX{{\rm B\kern-.05em{\sc i\kern-.025em b}\kern-.08em
    T\kern-.1667em\lower.7ex\hbox{E}\kern-.125emX}}
   
\begin{document}

\title{Enhancing 3GPP Urban Channel Models For Terrestrial-to-Non-Terrestrial Communication\\
\thanks{This work has been submitted to the IEEE for possible publication.
Copyright may be transferred without notice, after which this version may
no longer be accessible.}
}

\author{\IEEEauthorblockN{Gerhard Schreiber\textsuperscript{*}, Chenrui Sun\textsuperscript{§}, Joerg Schaepperle\textsuperscript{*}, Le-Hang Nguyen\textsuperscript{*}, Thorsten Wild\textsuperscript{*}}
\IEEEauthorblockA{\textit{\textsuperscript{*}Nokia Bell Labs, Radio Systems Research, Stuttgart, Germany } \\
\textit{\textsuperscript{§}School of Physics, Engineering and Technology, University of York, United Kingdom}\\
\{gerhard.schreiber, joerg.schaepperle, le\_hang.nguyen, thorsten.wild\}@nokia-bell-labs.com, chenrui.sun@york.ac.uk}
}

\maketitle

\begin{abstract}
We have developed enhanced models for large-scale channel parameters intended for terrestrial-to-non-terrestrial (NTN) communication in Urban environments for supporting devices above ground levels, such as drones. The models are formulated as functions of terminal height above ground and elevation angle, applicable to S-Band and Ka-Band frequencies. To ensure realistic models, an open-source 3D scene generator was utilized to create a diverse set of urban scenes across Europe. Sionna ray-tracing was employed to generate path-gain samples for outdoor terminals, covering a range of heights up to 300 meters and elevation angles up to 90°. Our proposed analytical functions show excellent agreement to ray-tracing data for line-of-sight (LOS) probability, shadow-fading (SF), and clutter-loss (CL). Furthermore, a comparative coupling gain analysis with existing 3GPP models reveals notable differences, particularly caused by smaller CL and higher LOS probabilities from proposed models. Designed for ease of use, these models can be seamlessly integrated into simulation tools, offering a practical solution for researchers and engineers.
\end{abstract}


\section{Introduction and Problem Statement}

\input{01_Introduction}

\section{Methodology}

\input{02_Methodology}

\section{Modeling large-Scale Parameters}

\input{03_LineOfSightProbabilityResults}

\input{04_ShadowFadingModelResults}

\input{05_ClutterLossModelResults}

\section{Comparison to 3GPP Models}

\input{07_CouplingGainComparison}

\section{Conclusion}

\input{08_Conclusion}

\section*{Acknowledgment}

\input{09_Acknowlegement}

\bibliographystyle{./IEEEtran}
\bibliography{./IEEEabrv,./bibliography}

\end{document}

%% file: 01_Introduction.tex
Unmanned aerial vehicles (UAV) have become indispensable across a broad spectrum of industries and non-commercial use cases \cite{11503149}. The viability and success of these use cases are based on fulfilling a set of stringent requirements including maintaining both communication and operational reliability. To meet these requirements, wireless networks with aerials must provide ultra-reliable communication and low-latency connectivity to ensure seamless integration and compliance with regulatory standards.
The combination of terrestrial (TN) and non-terrestrial (NTN) networks for connecting aerials offers an attractive option to further enhance the connectivity quality. To assess the potential benefits of this combination appropriate channel models are needed for network planning and system-simulation tools. An evaluation revealed that existing NTN channel models from \cite{3GPP38.811} are valid for terminals on ground and show inconsistencies in SF and CL parameterization. Furthermore, 3GPP creation methodology had, to our best knowledge, not been disclosed. Therefore, current 3GPP models are not applicable to accurately evaluate connections of an aerial terminal to a high altitude basestation.

The main contribution of this paper is to provide detailed insight on how we (i) used public tools to generate 3D-Urban scenarios and (ii) used the Sionna ray tracer to obtain large sets of channel data to (iii) derive analytic expressions for LOS probability, SF deviation and CL as function of terminal height and elevation angle towards a high altitude basestation.

%% file: 02_Methodology.tex
\subsection{Scene Generation and Ray-Tracing}\label{SCENE_GENERATION_RAY_TRACING}
\begin{figure}[htbp]
\centerline{\includegraphics[width=0.55\columnwidth]{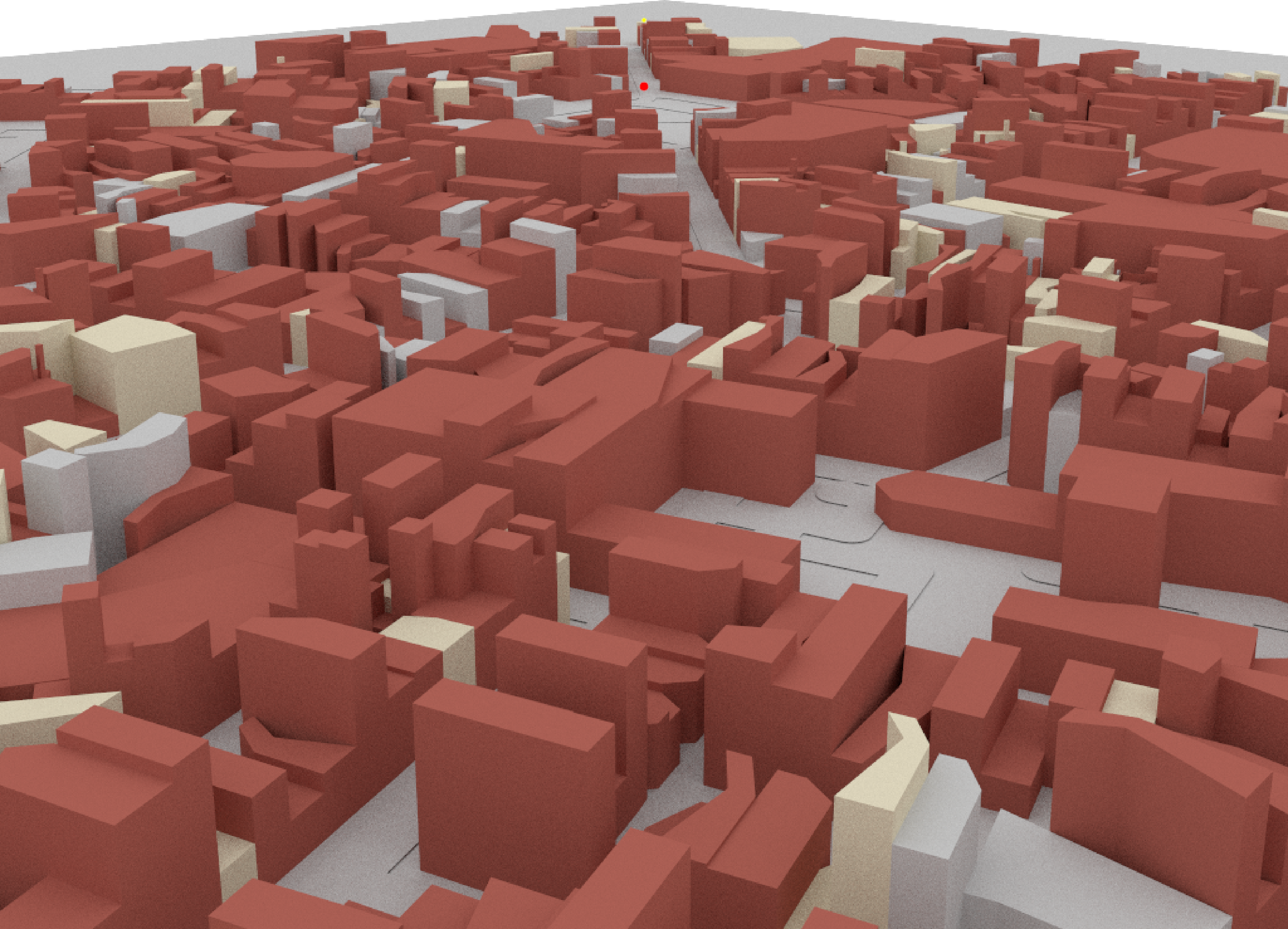}}
\caption{City of Lille as example for a generated Urban scene.}
\label{fig:urban_city}
\end{figure}

The recommendation \cite{ITU-R_P.2402} describes a prediction method to determine earth-space clutter loss from cumulative distributions of horizontal distances and building heights. An analysis of this approach revealed that it may not be the appropriate solution, due to inherent methodological limitations, like simplified statistical assumptions on building shapes and distribution and simple reflection loss modeling. In order to obtain more realistic large-scale channel parameter models, we developed a processing chain comprising two key modules that specialize in scene creation and ray tracing (Sionna).
The first module in the chain is an open-source 3D scene generator \cite{zumg}, responsible for generating building, street, and ground maps using the OpenStreetMap (OSM) \cite{osm} database for accurate spatial representation of the physical environment.
Building materials are randomly assigned (80\% brick-stone, 10\% concrete and 10\% marble) as OSM data does not provide this information.
The street and ground elements are modeled as concrete. Furthermore, building heights are limited to \mbox{22.5 m} to comply with maximum terminal height for 3GPP-Urban case. In total we created 20 Urban scenes from locations across Europe. Additional scenes can be artificially created by rotation of the high altitude basestation around the center of the scene; this allows breaking symmetries, e.g. caused by city planning.
Fig. \ref{fig:urban_city} shows an example for the city of Lille. 

The second module is the Sionna ray tracer \cite{sionna} that is used to determine the LOS/NLOS propagation conditions and the path-gains (PG) of aerial terminal towards a high altitude basestation. PGs are determined in horizontal meshes with \mbox{4 m} resolution and terminals are positioned at grid points. 
Channel parameter evolution is expected to show pronounced variability primarily in below-rooftop situations, where finer vertical spatial resolutions are required to capture the intricate changes. To address this, we employ a resolution of \mbox{$\sim$1 m}, ensuring high granularity in data collection. In contrast, for above-rooftop cases we expect more gradual variations across larger spatial scales. Consequently, relaxed resolutions in range \mbox{3-30 m} are applied to effectively capture these broader trends.
Different elevation angles are configured by corresponding positioning of the basestation. Tab. \ref{tab:SIONNA_RT_PARAMS_AND_SCENE_SETUP} summarizes the key assumptions to generate 3D scenes and to conduct ray tracing campaigns. The obtained results from the ray tracing are then used for subsequent statistical analysis to obtain analytic models for large-scale channel parameters.
\captionsetup{font=figure_size}
\captionsetup{font=normal_size}
\begin{table}[htbp]
\caption{Ray Tracing Configuration And Scenario Setup.}
\begin{center}
    \setlength{\tabcolsep}{4pt}
    \begin{tabular}{c|c}    
    \hline
    Parameter & Value  \\
    \hline
    Carrier frequency & 2 GHz (S-Band), 28 GHz (Ka-Band) \\
    \hline  
    Scenario & Urban \\
    \hline  
    Area dimension & 0.75 km x 0.75 km \\
    \hline  
    Number locations $N_{loc}$ & 20 across Europe \\
    \hline  
    Specular reflection & Disabled \\
    \hline  
    Diffuse reflection & Enabled, scattering coefficient: 0.3 \\
    & directivity parameter: 10 \\    
    \hline  
    Diffraction & Enabled \\
    \hline  
    Refraction & Disabled \\
    \hline  
    Max. number ray interactions & 4 \\
    \hline  
    Number rays & 3E9  \\
    \hline  
    BS antenna & X-pol, isotropic \\
    \hline  
    BS height & 10 km \\    
    \hline  
    BS elevation angles $\Theta$ & [10, 20,..., 90]{\large°} \\  
    towards scene center & \\
    \hline  
    UT location & 100\% Outdoor \\
    \hline  
    UT antenna & X-pol, isotropic \\
    \hline  
    UT height & [1.5, 300] m \\
    \hline
    Grid size & 4 m x 4 m \\
    \hline
    Building materials & random assignment, 80\% brick stone \\
                       & 10\% concrete, 10\% marble\\
    \hline
    Ground and Road material & concrete \\                
    \hline                    
    Max. building height $h_{RT}$ & 22.5 m \\
    \hline   
    \end{tabular}
    \label{tab:SIONNA_RT_PARAMS_AND_SCENE_SETUP}
\end{center}    
\end{table}

\subsection{Channel Data Analytics}\label{CHANNEL_DATA_ANALYTICS}
Each terminal position is classified as either line-of-sight (LOS)
or non-line-of-sight (NLOS). The traditional definition of LOS is employed namely, that the terminals have an unobstructed ray to the basestation. From the raw data we determine the per location LOS probability as ratio between the number terminals in LOS and total number terminals as function of terminal height $h$ above ground and elevation  $\theta$. The final result is the average over all locations:
\noindent
\begin{equation}
{P}_{LOS,raw}({h},\theta) = \frac{1}{N_{loc}}\mathlarger{\sum}_{loc=1}^{N_{loc}}\left(\frac{N_{LOS}(h,\alpha\mapsto\theta)}{N_{total}(h,\alpha\mapsto\theta)}\right)_{loc},
\label{eqn:p_los_raw}
\end{equation}
\noindent
where a mapping of the actual elevation angle $\alpha$ to the nearest, 3GPP compliant, grid elevation angle $\theta$ is employed: 
\noindent
\begin{equation}
\theta =\underset{\theta'\in \Theta}{\text{min}}\,|\alpha-\theta'|,
\label{eqn:tetha}
\end{equation}
\noindent
where $\Theta = \{10,20,30,40,50,60,70,80,90\}${\large°}. Typically the distance of angle $\alpha$ is within $\pm 2${\large°} from a grid elevation angle, so model accuracy in respect to elevation is preserved.
In general the total path gain $PG$ for given distance $d$, terminal height and an elevation $el$ is 
calculated using the following formula:
\noindent
\begin{equation}
PG(d,h,el) = FSPG(d)-CL(h,el)+N(0,SF,h,el),
\label{eqn:pg}
\end{equation}
\noindent
with $FSPG$ the Friis \cite{friis} free space path gain, as a function of the link distance, $d$ [m], and
carrier frequency, $f$ [Hz]:
\noindent
\begin{equation}
FSPG = 20\log_{10}\left(c_0/(4\pi fd)\right)  \text{[dB]},
\label{eqn:FSPG}
\end{equation}
\noindent
$CL$ is the clutter-loss and $N$ the random shadowing. 
According to \cite{ITU-R_P.1406} shadow fading refers to the slow, large-scale signal fluctuations caused by environmental obstacles (like buildings, terrain, and trees). It is mathematically treated as a zero-mean normal distribution with a standard deviation SF. From our ray-tracing simulations the SF is therefore determined as standard deviation over all measured path gains $PG$ as function of  elevation and height:
\noindent
\begin{equation}
{SF}_{raw}({h},\theta) = \frac{1}{N_{loc}}\sum_{loc=1}^{N_{loc}}{\text{std}(PG(h,\alpha\mapsto\theta))_{loc}},
\label{eqn:sf_raw}
\end{equation}
\noindent
where the $FSPG$ contribution has been omitted, because it evaluates to almost the same value for all considered terminals. 
For the CL we follow the recommendation \cite{ITU-R_P.2108} and \cite{3GPP38.811} and 
interpret this contribution as an additive correction to the free space path propagation model. 
Clutter and clutter-loss are defined in \cite{ITU-R_P.2108} and according to this
recommendation clutter originates from objects on ground that cannot be considered as pure terrain, and examples are buildings, vegetation, vehicles etc. In an Urban environment the
clutter  strongly depends on the surface materials, the building shape and on how 
buildings are distributed over the considered area. The clutter-loss is calculated as 
median over all differences between the $FSPGs$ and measured $PGs$, i.e. effectively the difference in path-gains without and with clutter over the communication link, as function of height and elevation:
\noindent
{
\begin{IEEEeqnarray}{rll}
\setlength{\nulldelimiterspace}{0pt}
& {CL}_{raw}({h},\theta) = \nonumber \\
& = \frac{1}{N_{loc}}\sum_{loc=1}^{N_{loc}}{\text{median}(FSPG-PG(h,\alpha\mapsto\theta))_{loc}}.
\end{IEEEeqnarray}
}
\noindent
CL might be negative due to constructive summation of multiple rays. All $SF_{raw}$ and $CL_{raw}$ numbers are calculated individually for LOS and NLOS propagation condition. The $SF_{raw}({h},\theta)$ and $CL_{raw}({h},\theta)$ sample data are then used to develop the analytic models.

%% file: 03_LineOfSightProbabilityResults.tex
\subsection{Line-of-Sight Probability}
\begin{figure}[htbp]
\centerline{\includegraphics[width=0.75\columnwidth]{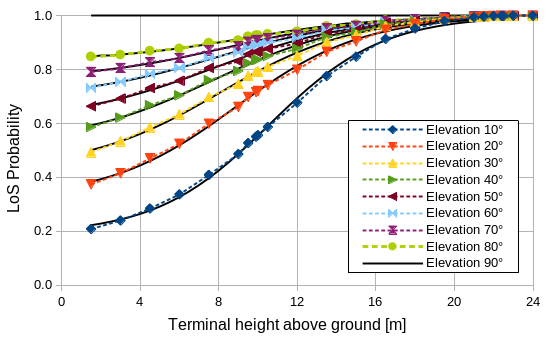}}
\caption{Comparison of raw data (marked colored lines) against model fit (solid black lines) of the LOS probability in S-Band.}
\label{fig:los_probability}
\end{figure}

For terminal heights below ($1.5\text{ m}\leq h\leq{h}_{RT}$) rooftop we found that the LOS probability can be closely approximated by:
\noindent
\begin{equation}
{P}_{LOS}({h},\theta) = S({h},A_\theta,1,X_\theta,W_\theta),
\label{eqn:below}
\end{equation}
\noindent
with $S$ being a Sigmoid-function of the following form:
\begin{equation}
S(H,A,B,X,W)=A\cdot\biggl( 1-\frac{1}{1+e^{-(H-X)/W}}\biggr)+B,
\label{eqn:sigmoid_pref}
\end{equation}
\noindent
which is similar to the expression from \cite{hourani}.
The parameters $A_\theta,X_\theta,W_\theta$ are provided in Table \ref{tab:LOS_PROBABILITY}, For terminals above rooftop ($h>{h}_{RT}$) or at 90{\large°} elevation the LOS probability is set to value 1. Fig. \ref{fig:los_probability} shows the raw and modeled the LOS probability as function of elevation angle and terminal height above ground.
\noindent
\input{TABLE_LineOfSightProbability}
\noindent
\captionsetup{font=figure_size}
\noindent

%% file: TABLE_LineOfSightProbability.tex
\captionsetup{font=normal_size}
\begin{table}[htbp]
\caption{LOS probability model parameters for $1.5\text{ m}\leq h\leq{h}_{RT}$.}
\begin{center} 
    \setlength{\tabcolsep}{2pt}
    \begin{tabular}{c|ccc|c|ccc|c|ccc}    
    \hline
    $\theta\text{\,}[\text{°}]$ & $A_\theta$ & $X_\theta$ & $W_\theta$ & $\theta$ & $A_\theta$ & $X_\theta$ & $W_\theta$ & $\theta$ & $A_\theta$ & $X_\theta$ & $W_\theta$ \\ 
    \hline
    10	&   -0.811		&	10.553	&	2.871   & 40    &	-0.476	&	7.745	&	3.505 & 70	&	-0.235	&	8.422	&	3.538 \\
    20	&	-0.679		&	8.847	&	3.218	& 50	&	-0.390	&	7.701	&	3.581 & 80	&	-0.167	&	9.305	&	3.450 \\
    30	&	-0.571		&	8.054	&	3.394	& 60	&	-0.308	&	7.890	&	3.580 & 90	&	0.000	&	0.000	&	1.000 \\
    \hline 
    \end{tabular}
    \label{tab:LOS_PROBABILITY}
\end{center}    
\end{table}

%% file: 04_ShadowFadingModelResults.tex
\subsection{Shadow-Fading}
\begin{figure}[htbp]
\centerline{\includegraphics[width=0.75\columnwidth]{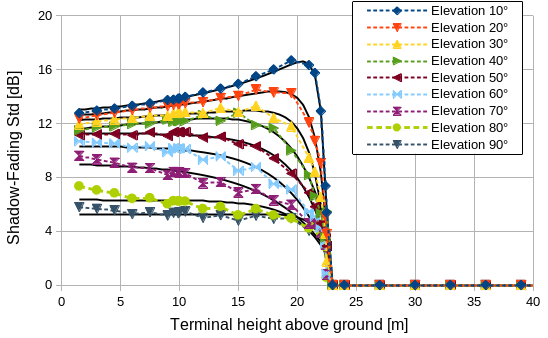}}
\caption{Comparison of raw data (marked colored lines) against model fit (solid black lines) for shadow-fading under NLOS propagation condition in S-Band.}
\label{fig:sf_nlos}
\end{figure}

For the SF quantity we have to differentiate between LOS and NLOS propagation condition and whether the terminal is below or above rooftop. For LOS propagation condition and below rooftop we found that this quantity is best modeled by:
\noindent
\begin{equation}
{SF}_{LOS}(h,\theta) = S(h,A_\theta,B_\theta,X_\theta,W_\theta),
\label{eqn:sf_los_below}
\end{equation}
\noindent
with parameters given in Table \ref{tab:SF_LOS_BELOW_RT}. For above rooftop (${h}_{RT}<{h}\leq 300\text{ m}$) the quantity is best represented by:
\noindent
\begin{equation}
{SF}_{LOS}(h,\theta) = M(\Delta{h},A_{1,\theta},W_{1,\theta},\Delta{h},A_{2,\theta},W_{2,\theta}),
\label{eqn:sf_los_above}
\end{equation}
\noindent
with $M$ a Morse-potential \cite{morse} like function as follows:
\begin{equation}
M(H_1,A_1,W_1,H_2,A_2,W_2) = A_1\cdot e^{-W_1H_1}-A_2\cdot e^{-W_2H_2},
\label{eqn:morse}
\end{equation}
\noindent
and $\Delta{h}$ being the difference between the terminal and maximum rooftop height:
\noindent
\begin{equation}
\Delta{h} = {h}-h_{RT} \text{ [m]}.
\label{eqn:h_UT}
\end{equation}
\noindent
The parameters $A_{1,\theta},W_{1,\theta},A_{2,\theta},W_{2,\theta}$ are provided in Table \ref{tab:SF_LOS_ABOVE_RT}.
\captionsetup{font=figure_size}
\begin{figure}[htbp]
\centerline{\includegraphics[width=0.75\columnwidth]{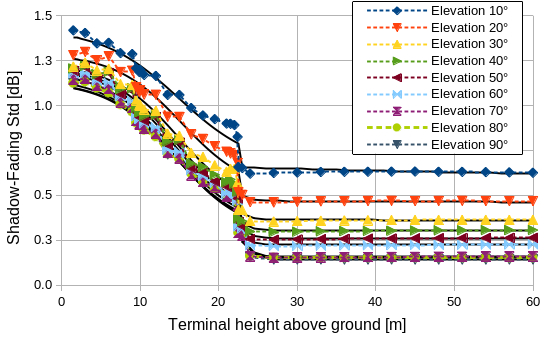}}
\caption{Comparison of raw data (marked colored lines) against model fit (solid black lines) for shadow-fading under LOS propagation condition  environment in S-Band.}
\label{fig:sf_los}
\end{figure}
As shown in Fig. \ref{fig:sf_los}, we obtained a good match between raw and modeled SF values. For above rooftop SF values show a significant positive value, that is caused by back-reflected rays. In case NLOS propagation condition and below rooftop the SF quantity is best modeled by:
\noindent
\begin{equation}
{SF}_{NLOS}({h},\theta) = M({h},A_{1,\theta},W_{1,\theta},\Delta{h},A_{2,\theta},W_{2,\theta})+A_{2,\theta},
\label{eqn:sf_nlos_below}
\end{equation}
\noindent
with parameters provided in Table \ref{tab:SF_NLOS_BELOW_RT}. For terminals above rooftop (${h}>{h}_{RT}$) the quantity is equal 0. Fig. \ref{fig:sf_nlos} shows a comparison between raw and modeled results.
\noindent
\captionsetup{font=figure_size}
\noindent
\input{TABLE_ShadowFading}

%% file: TABLE_ShadowFading.tex
\noindent
\captionsetup{font=normal_size}
\begin{table}[htbp]
\caption{Shadow-Fading model parameters for $1.5\text{ m} \leq{h} \leq {h}_{RT}$ under LOS propagation condition.}
\begin{center}
    \setlength{\tabcolsep}{2pt}
    \begin{tabular}{c|cccc|cccc}    
    \hline
    & \multicolumn{4}{c|}{S-Band} &  \multicolumn{4}{c}{Ka-Band} \\
    \hline
    $\theta\text{\,}[\text{°}]$ & $A_\theta$ & $B_\theta$ & $X_\theta$ & $W_\theta$ & $A_\theta$ & $B_\theta$ & $X_\theta$ & $W_\theta$  \\ 
    \hline
    10	&	0.771	&	0.655	&	14.915	&	4.833	&	0.887	&	0.619	&	13.242	&	3.929	\\
    20	&	0.774	&	0.516	&	15.271	&	4.236	&	0.866	&	0.481	&	13.610	&	3.607	\\
    30	&	0.810	&	0.418	&	14.881	&	4.128	&	0.904	&	0.378	&	13.572	&	3.625	\\
    40	&	0.847	&	0.370	&	14.548	&	4.108	&	0.939	&	0.328	&	13.464	&	3.623	\\
    50	&	0.853	&	0.351	&	14.374	&	4.071	&	0.953	&	0.297	&	13.548	&	3.646	\\
    60	&	0.864	&	0.326	&	14.180	&	4.193	&	0.965	&	0.258	&	13.484	&	3.754	\\
    70	&	0.866	&	0.289	&	14.348	&	4.239	&	0.966	&	0.211	&	13.667	&	3.811	\\
    80	&	0.844	&	0.291	&	14.460	&	4.282	&	0.951	&	0.211	&	13.710	&	3.752	\\
    90	&	0.823	&	0.312	&	14.291	&	4.163	&	0.941	&	0.209	&	13.817	&	3.722	\\
    \hline 
    \end{tabular}
    \label{tab:SF_LOS_BELOW_RT}
\end{center}    
\end{table}
\noindent
\captionsetup{font=normal_size}
\begin{table}[htbp]
\caption{Shadow-Fading model parameters for ${h}_{RT} < {h} \leq 300\text{ m}$ under LOS propagation condition.}
\begin{center}
    \setlength{\tabcolsep}{2pt}
    \begin{tabular}{c|cccc|cccc}    
    \hline
    & \multicolumn{4}{c|}{S-Band} &  \multicolumn{4}{c}{Ka-Band} \\
    \hline    
    $\theta\text{\,}[\text{°}]$ & $A_{1,\theta}$ & $W_{1,\theta}$ & $A_{2,\theta}$ & $W_{2,\theta}$ & $A_{1,\theta}$ & $W_{1,\theta}$ & $A_{2,\theta}$ & $W_{2,\theta}$ \\
    \hline
    10	&	9.14E-2	&	1.46E-2	&	-0.565	&	-1.19E-4	&	0.550	&	1.19E-3	&	-7.06E-2	&	-3.65E-3	\\
    20	&	4.75E-2	&	1.00E+0	&	-0.469	&	4.30E-4	&	0.549	&	6.23E-4	&	6.68E-2	&	6.23E-4	\\
    30	&	5.16E-2	&	1.00E+0	&	-0.366	&	5.76E-4	&	0.549	&	8.03E-4	&	1.70E-1	&	8.03E-4	\\
    40	&	6.39E-2	&	1.00E+0	&	-0.307	&	3.31E-4	&	0.549	&	8.23E-4	&	2.19E-1	&	8.23E-4	\\
    50	&	9.01E-2	&	1.00E+0	&	-0.262	&	1.26E-4	&	0.549	&	1.06E-3	&	2.50E-1	&	1.06E-3	\\
    60	&	1.03E-1	&	1.00E+0	&	-0.223	&	-1.71E-4	&	0.549	&	8.58E-4	&	2.89E-1	&	8.58E-4	\\
    70	&	1.33E-1	&	9.35E-1	&	-0.156	&	-2.88E-4	&	0.538	&	1.74E-3	&	3.25E-1	&	1.74E-3	\\
    80	&	1.39E-1	&	7.67E-1	&	-0.153	&	1.62E-4	&	0.538	&	2.84E-3	&	3.25E-1	&	2.84E-3	\\
    90	&	1.71E-1	&	7.76E-1	&	-0.142	&	1.37E-4	&	0.538	&	3.61E-3	&	3.27E-1	&	3.61E-3	\\
    \hline 
    \end{tabular}
    \label{tab:SF_LOS_ABOVE_RT}
\end{center}    
\end{table}
\noindent
\captionsetup{font=normal_size}
\begin{table}[htbp]
\caption{Shadow-Fading model parameters for $1.5\text{ m} \leq{h} \leq {h}_{RT}$ under NLOS propagation condition.}
\begin{center}
    \setlength{\tabcolsep}{2pt}
    \begin{tabular}{c|cccc|cccc}    
    \hline
    & \multicolumn{4}{c|}{S-Band} &  \multicolumn{4}{c}{Ka-Band} \\
    \hline
    $\theta\text{\,}[\text{°}]$ & $A_{1,\theta}$ & $W_{1,\theta}$ & $A_{2,\theta}$ & $W_{2,\theta}$ & $A_{1,\theta}$ & $W_{1,\theta}$ & $A_{2,\theta}$ & $W_{2,\theta}$ \\
    \hline 
    10	&	5.28E-1	&	-1.04E-1	&	12.393	&	-1.927	&	0.711	&	-9.81E-2	&	15.821	&	-1.951	\\
    20	&	1.18E+0	&	-5.50E-2	&	11.334	&	-0.975	&	0.878	&	-7.28E-2	&	15.511	&	-0.816	\\
    30	&	1.23E+0	&	-3.27E-2	&	10.966	&	-0.749	&	0.829	&	-5.80E-2	&	15.055	&	-0.560	\\
    40	&	9.45E-1	&	-6.63E-2	&	10.584	&	-0.334	&	1.160	&	-6.04E-2	&	14.057	&	-0.319	\\
    50	&	1.06E-3	&	-3.56E-1	&	11.265	&	-0.356	&	2.471	&	-1.21E-2	&	12.147	&	-0.271	\\
    60	&	6.17E-3	&	-2.67E-1	&	10.336	&	-0.267	&	5.619	&	3.86E-2	&	8.910	&	-0.322	\\
    70	&	4.06E-2	&	-1.86E-1	&	9.078	&	-0.186	&	7.498	&	5.60E-2	&	5.760	&	-0.451	\\
    80	&	3.79E-4	&	-3.86E-1	&	6.323	&	-0.386	&	5.804	&	6.00E-2	&	3.820	&	-1.286	\\
    90	&	6.84E-18	&	-1.82E+0	&	5.265	&	-1.183	&	3.340	&	5.31E-2	&	3.752	&	-9.332	\\
    \hline 
    \end{tabular}
    \label{tab:SF_NLOS_BELOW_RT}
\end{center}    
\end{table}

%% file: 05_ClutterLossModelResults.tex
\subsection{Clutter-Loss}
\begin{figure}[htbp]
\centerline{\includegraphics[width=0.75\columnwidth]{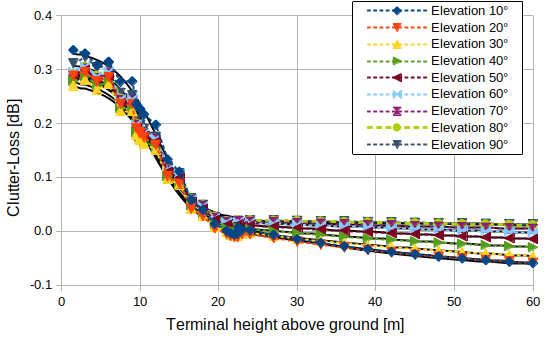}}
\caption{Comparison of raw data (marked colored lines) against model fit (solid black lines) for clutter-loss under LOS propagation condition in S-Band.}
\label{fig:cl_los}
\end{figure}

\begin{figure}[htbp]
\centerline{\includegraphics[width=0.75\columnwidth]{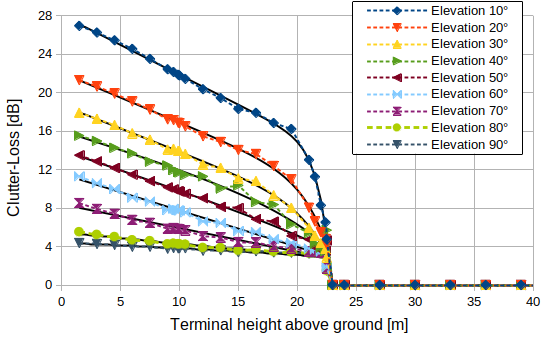}}
\caption{Comparison of raw data (marked colored lines) against model fit (solid black lines) for clutter-loss under NLOS propagation condition in S-Band.}
\label{fig:cl_nlos}
\end{figure}

As in case for SF we differentiate between LOS and NLOS propagation condition and whether the terminal is below or above rooftop. For LOS propagation condition and below rooftop the CL quantity an accurate approximation is provided by the following expression:
\noindent
\begin{equation}
{CL}_{LOS}({h},\theta) = S({h},A_\theta,B_\theta,X_\theta,W_\theta),
\label{eqn:cl_los_below}
\end{equation}
\noindent
with parameters given in Table \ref{tab:CL_LOS_BELOW_RT} and for above rooftop (${h}_{RT}<{h}\leq 300\text{ m}$) by:
\noindent
\begin{equation}
{CL}_{LOS}({h},\theta) = M(\Delta{h},A_{1,\theta},W_{1,\theta},\Delta{h},A_{2,\theta},W_{2,\theta}),
\label{eqn:cl_los_above}
\end{equation}
\noindent
using parameters provided in Table \ref{tab:CL_LOS_ABOVE_RT}. From Fig. \ref{fig:cl_los} we observe that the LOS CL is small for any elevation angle and terminal height. Therefore, one may set the LOS CL equal to value 0, an assumption that would correspond to current 3GPP model \cite{3GPP38.811}. A value unequal zero is equivalent to a deviation from the ideal exponent 2 in FSPG gain model. For terminal heights below rooftop the LOS CL is positive, whereas for location above rooftop the loss is slightly negative, indicating that back-reflected rays contributed to some positive PG.
\noindent
\captionsetup{font=figure_size}
\noindent
In case NLOS propagation and below rooftop the CL quantity is best modeled by  combining a linear and an exponential-function:
\noindent
\begin{equation}
{CL}_{NLOS}({h},\theta) = A_\theta-B_\theta\cdot{h} - C_\theta\cdot e^{-W_\theta\Delta{h}},
\label{eqn:cl_nlos_below}
\end{equation}
with parameters provided in Table \ref{tab:CL_NLOS_BELOW_RT}. For terminals above rooftop  the quantity is equal 0 as shown in Fig. \ref{fig:cl_nlos}.
\noindent
\captionsetup{font=figure_size}
\noindent
\input{TABLE_CluterLoss}

%% file: TABLE_CluterLoss.tex
\noindent
\captionsetup{font=normal_size}
\begin{table}[htbp]
\caption{Clutter-Loss model parameters for $1.5\text{ m} \leq{h} \leq {h}_{RT}$ under LOS propagation condition.}
\begin{center}
    \setlength{\tabcolsep}{2pt}
    \begin{tabular}{c|cccc|cccc}    
    \hline
    & \multicolumn{4}{c|}{S-Band} &  \multicolumn{4}{c}{Ka-Band} \\
    \hline
    $\theta\text{\,}[\text{°}]$ & $A_\theta$ & $B_\theta$ & $X_\theta$ & $W_\theta$ & $A_\theta$ & $B_\theta$ & $X_\theta$ & $W_\theta$  \\ 
    \hline 
    10	&	0.331	&	5.04E-3	&	12.273	&	2.734	&	0.456	&	1.74E-2	&	11.918	&	2.899	\\
    20	&	0.294	&	-2.36E-3	&	11.948	&	2.704	&	0.399	&	1.39E-3	&	11.546	&	2.903	\\
    30	&	0.274	&	-3.15E-4	&	11.956	&	2.706	&	0.358	&	3.15E-3	&	11.470	&	2.908	\\
    40	&	0.277	&	5.71E-3	&	11.823	&	2.744	&	0.352	&	8.48E-3	&	11.316	&	2.937	\\
    50	&	0.280	&	1.20E-2	&	11.873	&	2.738	&	0.342	&	1.46E-2	&	11.349	&	2.930	\\
    60	&	0.283	&	1.64E-2	&	11.871	&	2.737	&	0.330	&	1.83E-2	&	11.440	&	2.923	\\
    70	&	0.275	&	1.72E-2	&	11.980	&	2.680	&	0.301	&	1.81E-2	&	11.737	&	2.850	\\
    80	&	0.276	&	1.93E-2	&	11.904	&	2.663	&	0.307	&	2.17E-2	&	11.604	&	2.828	\\
    90	&	0.296	&	1.89E-2	&	11.723	&	2.719	&	0.311	&	2.23E-2	&	11.597	&	2.830	\\
    \hline 
    \end{tabular}
    \label{tab:CL_LOS_BELOW_RT}
\end{center}    
\end{table}
\noindent
\captionsetup{font=normal_size}
\begin{table}[htbp]
\caption{Clutter-Loss model parameters for ${h}_{RT} < {h} \leq 300\text{ m}$ under LOS propagation condition.}
\begin{center}  
    \setlength{\tabcolsep}{1.3pt}
    \begin{tabular}{c|cccc|cccc}    
    \hline
    & \multicolumn{4}{c|}{S-Band} &  \multicolumn{4}{c}{Ka-Band} \\
    \hline    
    $\theta\text{\!\,}[\text{°}]$ & $A_{1,\theta}$ & $W_{1,\theta}$ & $A_{2,\theta}$ & $W_{2,\theta}$ & $A_{1,\theta}$ & $W_{1,\theta}$ & $A_{2,\theta}$ & $W_{2,\theta}$ \\
    \hline 
    10	&	2.00E+0	&	1.85E-2	&	1.99E+0	&	1.68E-2	&	2.00E+0	&	2.13E-2	&	1.98E+0	&	2.00E-2	\\
    20	&	4.05E-1	&	1.39E-2	&	4.06E-1	&	8.09E-3	&	2.63E-1	&	1.64E-2	&	2.61E-1	&	7.11E-3	\\
    30	&	1.15E-1	&	1.81E-2	&	1.14E-1	&	2.31E-3	&	1.04E-1	&	2.10E-2	&	1.01E-1	&	2.31E-3	\\
    40	&	8.29E-2	&	1.62E-2	&	7.65E-2	&	5.23E-4	&	8.02E-2	&	1.89E-2	&	7.11E-2	&	8.67E-4	\\
    50	&	6.52E-2	&	1.37E-2	&	5.25E-2	&	-5.33E-4	&	6.59E-2	&	1.63E-2	&	5.07E-2	&	-2.09E-4	\\
    60	&	5.45E-2	&	1.10E-2	&	3.74E-2	&	-9.18E-4	&	5.82E-2	&	1.34E-2	&	3.93E-2	&	-1.05E-3	\\
    70	&	4.97E-2	&	8.01E-3	&	3.18E-2	&	-4.51E-4	&	5.47E-2	&	1.11E-2	&	3.59E-2	&	-1.33E-3	\\
    80	&	1.95E+0	&	3.20E-3	&	1.93E+0	&	3.09E-3	&	4.83E-2	&	8.09E-3	&	2.61E-2	&	-8.28E-4	\\
    90	&	1.97E+0	&	2.92E-3	&	1.96E+0	&	2.84E-3	&	5.27E-2	&	6.51E-3	&	2.99E-2	&	-1.86E-4	\\
    \hline 
    \end{tabular}
    \label{tab:CL_LOS_ABOVE_RT}
\end{center}    
\end{table}
\noindent
\captionsetup{font=normal_size}
\begin{table}[htbp]
\caption{Clutter-Loss model parameters for $1.5\text{ m} \leq{h} \leq {h}_{RT}$ under NLOS propagation condition.}
\begin{center} 
    \setlength{\tabcolsep}{2pt}
    \begin{tabular}{c|cccc|cccc}    
    \hline
    & \multicolumn{4}{c|}{S-Band} &  \multicolumn{4}{c}{Ka-Band} \\
    \hline    
    $\theta\text{\,}[\text{°}]$ & $A_\theta$ & $B_\theta$ & $C_\theta$ & $W_\theta$ & $A_\theta$ & $B_\theta$ & $C_\theta$ & $W_\theta$  \\ 
    \hline 
    10	&	28.014	&	0.616	&	9.151	&	-9.78E-1	&	35.638	&	0.713	&	14.812	&	-8.40E-1	\\
    20	&	22.069	&	0.516	&	6.341	&	-4.94E-1	&	28.077	&	0.614	&	10.371	&	-4.14E-1	\\
    30	&	18.675	&	0.480	&	4.744	&	-4.24E-1	&	23.337	&	0.556	&	7.783	&	-3.08E-1	\\
    40	&	16.076	&	0.380	&	3.464	&	-1.88E-1	&	20.365	&	0.426	&	6.756	&	-1.39E-1	\\
    50	&	14.046	&	0.412	&	1.000	&	-1.69E-1	&	17.658	&	0.543	&	1.815	&	-1.47E-1	\\
    60	&	13.382	&	0.379	&	1.824	&	-1.28E-6	&	16.153	&	0.506	&	1.874	&	2.86E-6	\\
    70	&	12.735	&	0.247	&	4.338	&	-1.36E-6	&	14.983	&	0.322	&	5.004	&	6.98E-7	\\
    80	&	10.190	&	0.117	&	4.716	&	3.15E-6	&	11.742	&	0.152	&	5.579	&	1.67E-6	\\
    90	&	40.000	&	0.280	&	30.737	&	6.56E-3	&	40.000	&	0.302	&	30.295	&	6.82E-3	\\
    \hline 
    \end{tabular}
    \label{tab:CL_NLOS_BELOW_RT}
\end{center}    
\end{table}

%% file: 07_CouplingGainComparison.tex
\begin{figure}[htbp]
\centerline{\includegraphics[width=0.75\columnwidth]{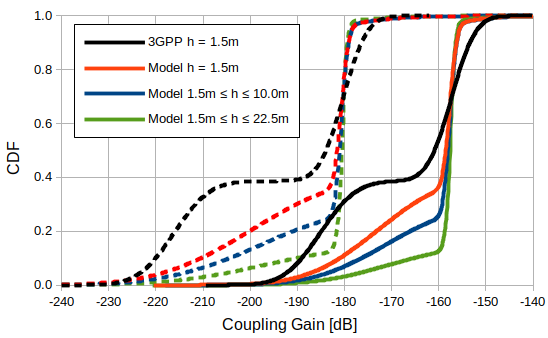}}
\caption{Coupling gains for 3GPP and proposed model assuming different terminal height distributions. The solid lines are for S-Band and the dashed lines are for Ka-Band.}
\label{fig:coupling_gain_urban}
\end{figure}

The downlink coupling gain cumulative distribution function (CDF) is widely employed as a key performance indicator in both open-source and proprietary link- and system-level simulators. This metric plays a pivotal role in validating the implementation of propagation models and conducting baseline radio coverage analyses, particularly within the scope of 3GPP study items. The coupling gain represents the net signal loss between a transmitter and receiver, encompassing contributions from path loss, antenna gains, clutter-loss, average shadowing, and, when applicable, fast fading.
To compare the performance of the current 3GPP model with our proposed model, we conducted an evaluation of coupling gain based on the assumptions detailed in Table \ref{tab:ASSUMPTION_COUPLING_GAIN}. The results, presented in Figure \ref{fig:coupling_gain_urban}, shows that the proposed model consistently delivers higher coupling gains across various carrier frequencies and terminal height distributions than the current 3GPP model. This outcome is primarily driven by two factors: Firstly, a substantial reduction in CL, which can exceed 20 dB in specific situations, and secondly an increased probability for LOS conditions at elevated terminal positions.

\begin{table}[htbp]
\caption{Coupling Gain Evaluation Assumptions.}
\begin{center} 
    \setlength{\tabcolsep}{4pt}
    \begin{tabular}{c|c}    
    \hline
    Parameter & Value  \\
    \hline
    Carrier frequency & 2 GHz (S-Band), 28 GHz (Ka-Band) \\
    \hline  
    Fast-Fading Channel &  not modeled \\
    \hline  
    Scenario & Urban \\
    \hline  
    Earth radius & 6371 km \\
    \hline
    Satellite position & latitude/longitude=(84.45{\large°}, 0.0{\large°}) \\
    \hline
    Satellite height & 600 km \\
    \hline
    Satellite antenna & X-pol, isotropic, 0 dBi gain \\
    \hline
    Cell center & latitude/longitude=(90.0{\large°}, 0.0{\large°}) \\
                & i. e. 40{\large°} elevation angle to satellite\\
    \hline
    UT coverage radius & 1000 m \\
    \hline
    UT positions & uniform on sphere within \\
                 & coverage radius around cell center\\
    \hline
    UT heights above ground & uniform \\
    \hline
    UT outdoor probability & 100\% \\    
    \hline
    UT elevation angle & 40{\large°} \\
    \hline
    UT antenna & X-pol, isotropic, 0 dBi gain \\
    \hline
    UT drops & 1E5 \\    
    \hline   
    \end{tabular}
    \label{tab:ASSUMPTION_COUPLING_GAIN}
\end{center}    
\end{table}

\captionsetup{font=figure_size}

%% file: 08_Conclusion.tex
In this work, we identified issues and limitations in the current 3GPP NTN channel models and emphasized the need to close these open research questions by provision of enhanced 3GPP compliant
models to accurately quantify the com-munication link between terrestrial/aerial and 
non-terrestrial devices. Using an open-source tool, we efficiently generated numerous 3D-Urban scenarios from OSM data and employed Sionna ray-tracing to analyze LOS probability and path gains under LOS and NLOS propagation conditions. This enabled the development of accurate analytic expressions for large-scale channel parameters like LOS probability, shadow fading, and clutter loss as function of terminal elevation angle and height above ground. Coupling gain evaluations revealed significant differences to existing 3GPP NTN models, primarily due to lower clutter-loss values and higher LOS probabilities of the proposed model. While initially designed for aerial terminals, the model may also be applicable to in-building terminals. Future work will extend the model creation to Dense Urban, Suburban, and Rural scenarios. Further advancements in large-scale channel parameter modeling can be achieved via refining 3D scene generation and improving ray-tracing accuracy.

%% file: 09_Acknowlegement.tex
This research was supported by the Federal Ministry of Research, Technology and Space (BMFTR) of Germany under grant 16KIS2424 (6G-Coverage).